# Monochromatic gamma emitter for low energy quanta

*Devoted to Vasil Gagov*

## Z. R. Tomova[1], D. Z. Zhechev[2] and S. A. Mironova[3]

[1]Institute of Nuclear Research and Nuclear Energy, Bulgarian Academy of Sciences
72 Tzarigradsko Shaussee Blvd., BG-1784 SOFIA, BULGARIA
e-mail:zoritzaclff@yahoo.com

[2]Institute of Solid State Physics, Bulgarian Academy of Sciences
72 Tzarigradsko Shaussee Blvd., BG-1784 SOFIA, BULGARIA
e-mail:spectron@issp.bas.bg

[3]Faculty of Physics, Sofia University
5 James Bourchier Blvd., BG-1164 SOFIA, BULGARIA

Abstract
The possibility of creating of a monochromatic gamma emitter of low energy quanta is analyzed. The idea is based on Daning's scheme. Except for purely scientific problems the monochromator is actual for therapy of wide range of diseases.

1. Introduction

A great number of applications of nuclear physical methods were realized with applications based on the monochromators of neutrons.
Here the possibility of creating a monochromatic gamma emitter of low energy quanta is analyzed. Except for purely scientific problems that monochromator is actual for therapy of wide range of diseases [1].
The monochromator suggested includes monochromator of neutrons, and the well-known scheme of Daning[2] is used.
The flight-method is known in the neutron spectroscopy:
The neutrons with fixed velocities are separated by using the difference in their flight time along the distance source -detector. For that purpose mechanical devices called mechanical selectors of neutrons are set on the way of the neutron flow.
In 1935 Daning built a monochromator with helical shape like slits. [2]. The modern mechanical monochromators for thermal and intermediate-energy $(1 - 1.10^{4}[eV])$ neutrons are based on the same principle. They contain a cylinder with curvilinear slits, longitudinal or cross to its axis of rotation. The rotator surface may contain a lot of helical slits so that the system outlet produces uninterrupted beam of monoenergetic neutrons. The energy resolution of the instrument depends on the rotator length, the angular speed of rotation and the width of its slits.

Namely, an instrument of that kind can be used for the excitation of a reaction(n,γ) of radiative capture of neutrons in layers from resonant absorbing materials: the energy of the excited nucleus is emitted in gamma quanta. Two boundary cases are possible: the final nucleus is a stable isotope of the irradiated one; the final nucleus is radioactive – the product are radioactive isotopes.

2. Principle and scheme realization of monochromatic gamma emitter for low energy quanta

The figures 1, 2, and 3 present the cross sections for radiative capture (*green color*) and the total cross section of Se-76, Fe-56 and Ni-60 (*red color*) vs the striking neutrons energy.

Plots are received from the internet site of **Korea Atomic Energy Research Institute**(KAERI)[3].

For Fe-56 and Ni-60 data from ENDF/B-6.1 [4], when for Se-76-from ENDF/B-VI, release 8 [5] is used.

Also, isotopes Cr-52, Ca-43, Fe-57, Mo-94, Nd-142, Sr-86, Xe-128, Xe-130, He-3 have similar properties.

The plots suggest that for certain energies of the initial neutron beam the process of radiative capture is resonant and the other possible processes are negligible.

The value of the energy, corresponding to the resonance maximum allows the calculation of the wavelength of the emitted monochromatic wave λ.

Our calculations show that for energies within the interval $10^3$-$10^5$ [eV] the emission is in the range of wavelengths typical for X- and gamma emission.

Furthermore, because of the resonant character of the reaction, X- and gamma emission is resonant and monochromatic with line width of the order of 0.2-0.3[eV] [2].

In addition, both initial and final products of the reaction are stable isotopes.

Fig. 4 illustrates a principle scheme of monochromatic gamma emitter based on the arguments discussed above.

It consists of:
1. a source of neutrons of intermediate energies;
2. fast mechanical separator tuned to the interval of energies characteristic for the next element–3 resonant absorption region.
3. layer absorbing neutrons of defined energy emitted from the element 2.
4. nozzle of truncated cone shape (for example from steel) collimating the light beam
5. protecting shell

In our opinion the suggested scheme should emit strictly monochromatic gamma quanta of low energies.

Therefore, a possible application of the scheme seems to be in the medicine for treatment of the diseases described in table 1[1], as well as for radiation therapy for some kinds of cancer [6,7,8,9,10,11,12,13,14].

For that purpose element 6 can be added to the elements from the fig. 4.

It represents cylindrical tube (for example from steel), which fixes the acted area size and by the element 6a chops the outgoing ray from element 4 to great number of lower intensity rays.

The same procedure is used in the therapeutical method Intensity Modulated Radiation Therapy (IMRT)[15], realized by the Nomos Corporation.

Distinctive marks of the proposed scheme are comparatively low energy of the outlet beam and the possibility to vary its diameter and emission energy.

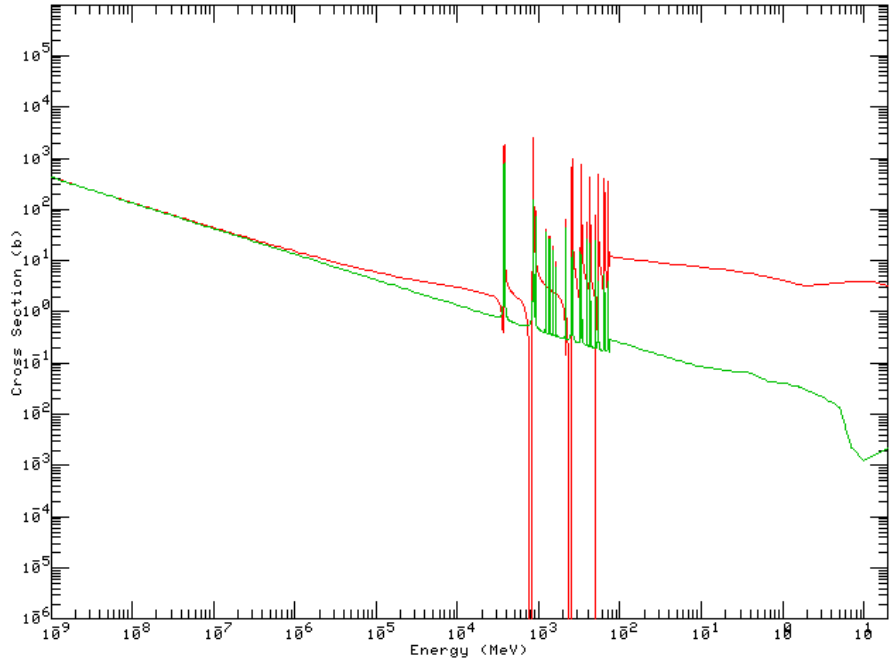

fig.1

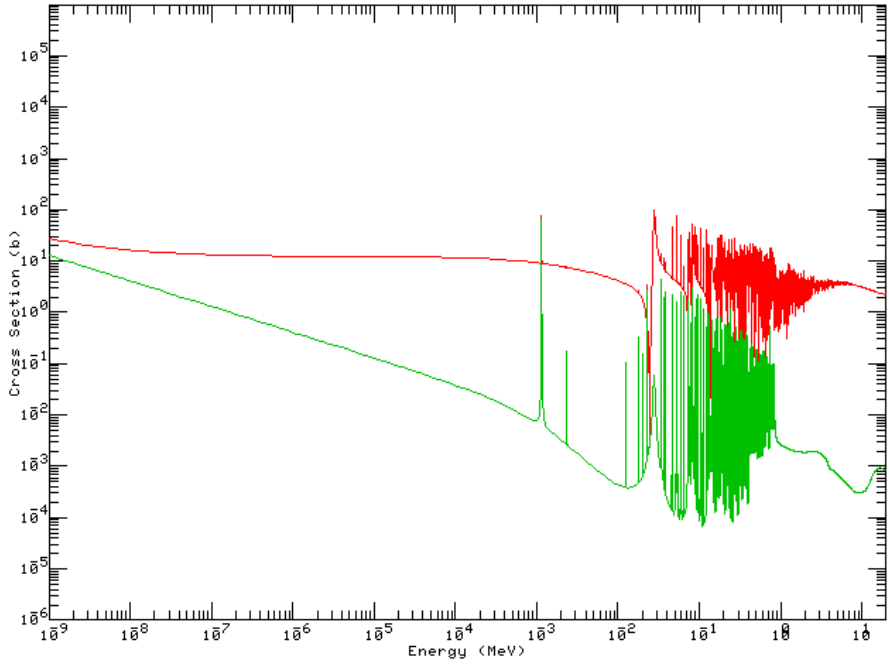

fig.2

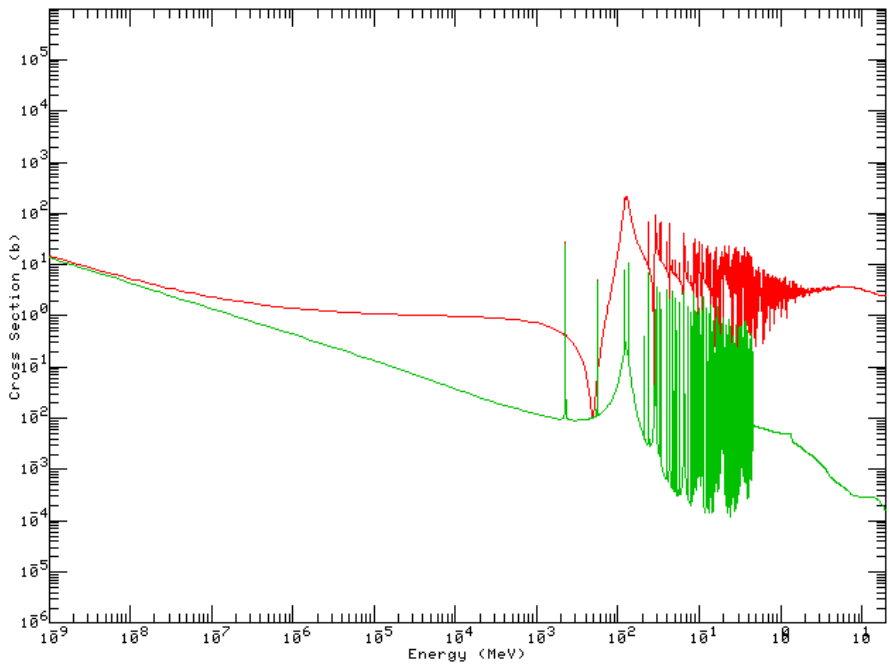

fig. 3

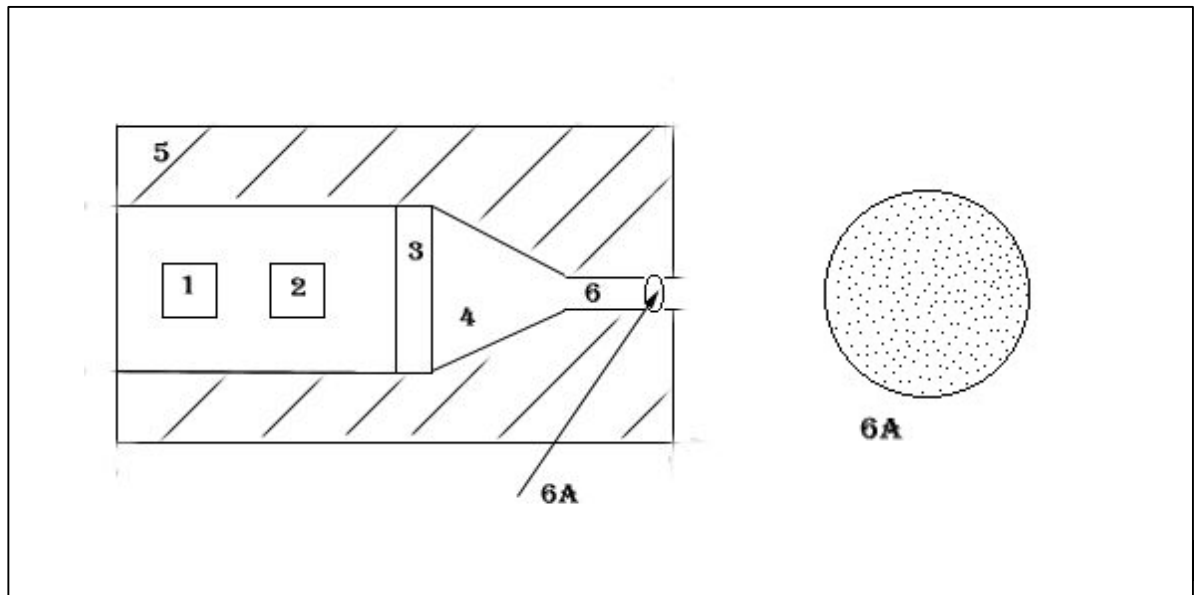

fig. 4

Table 1[1]

| Ameloblastoma |
|---|
| Aneurysmal bone cyst |
| Angiofibroma |
| Arteriovenous malformation |
| Chemodectoma |
| Chordoma |
| Craniopharyngioma |
| Desmoid tumor |
| Graves' ophthalmopathy |
| Gynecomastia associated with hormonal management of prostate cancer |
| Hemangioma |
| Heterotopic bone formation |
| Hypersplenism |
| Keloid |
| Keratoacanthoma |
| Meningioma |
| Peyronie's disease |
| Pituitary adenoma |
| Pterygium |
| Total lymphoid irradiation for autoimmune disease or organ transplantation |
| Vascular restenosis prevention |